\documentclass[12pt, english,journal, onecolumn, draftclsnofoot]{IEEEtran}
\usepackage[T1]{fontenc}
\usepackage[utf8]{inputenc}
\usepackage{prettyref}
\usepackage{refstyle}
\usepackage{amsbsy}
\usepackage{amsthm}
\usepackage{amssymb}
\usepackage{graphicx}
\usepackage{tikz}
\usepackage{pgfplots}
\usepackage{pgfgantt}

\pgfplotsset{compat=newest} 
\pgfplotsset{plot coordinates/math parser=false,             width=1.1\columnwidth,     
height=0.8\columnwidth}

\makeatletter

\AtBeginDocument{}
\AtBeginDocument{\providecommand\eqref[1]{\ref{eq:#1}}}
\AtBeginDocument{}
\AtBeginDocument{}
\RS@ifundefined{subsecref}
  {\newref{subsec}{name = \RSsectxt}}
  {}
\RS@ifundefined{thmref}
  {\def\RSthmtxt{theorem~}\newref{thm}{name = \RSthmtxt}}
  {}
\RS@ifundefined{lemref}
  {\def\RSlemtxt{lemma~}\newref{lem}{name = \RSlemtxt}}
  {}

\newcommand{\lyxaddress}[1]{
	\par {\raggedright #1
	\vspace{0.2em}
	\noindent\par}
}
\theoremstyle{definition}
\newtheorem{defn}{\protect\definitionname}
\theoremstyle{plain}
\newtheorem{lem}{\protect\lemmaname}
\theoremstyle{plain}

\theoremstyle{plain}
\newtheorem{thm}{\protect\theoremname}

\usepackage{hyperref}

\newref{eq}{name=Eq.~}
\newref{subsec}{name=Section~}
\newref{thm}{name=Theorem~}
\newref{lem}{name=Lemma~}
\newref{cor}{name=Corollary~}
\newref{fig}{name=Figure~}
\newref{def}{name=Definition~}

\makeatother

\usepackage{babel}
\providecommand{\corollaryname}{Corollary}
\providecommand{\definitionname}{Definition}
\providecommand{\lemmaname}{Lemma}
\providecommand{\theoremname}{Theorem}

\begin{document}
\title{Private Wireless Federated Learning with Anonymous Over-the-Air Computation}
\author{Burak Has\i rc\i o\u{g}lu and Deniz Gündüz}
\maketitle

\lyxaddress{\centering Information Processing and Communications Laboratory}
\lyxaddress{\centering Imperial College London, UK}
\lyxaddress{\centering \{b.hasircioglu18, d.gunduz\}@imperial.ac.uk}
\vspace{1.5em}

\begin{abstract}
In conventional federated learning (FL), differential privacy (DP) guarantees can be obtained by injecting additional noise to local model updates before transmitting to the parameter server (PS). In the wireless FL scenario, we show that the privacy of the system can be boosted by exploiting over-the-air computation (OAC) and anonymizing the transmitting devices. In OAC, devices transmit their model updates simultaneously and in an uncoded fashion, resulting in a much more efficient use of the available spectrum. We further exploit OAC to provide anonymity for the transmitting devices. The proposed approach improves the performance of private wireless FL by reducing the amount of noise that must be injected. \makeatletter{\renewcommand*{\@makefnmark}{}\footnotetext{This work was partially supported by the European Research Council (ERC) through project BEACON (No. 677854).}\makeatother}
\end{abstract}

\section{Introduction}

\label{sec:intro}

Federated learning (FL) \cite{mcmahan2017communication} allows multiple devices, each with its own local dataset, to train a model collaboratively with the help of a parameter server (PS) without sharing their datasets.  At each iteration of FL, the PS shares the most recent global model with the devices, which then use their local datasets to update the global model. Local updates are aggregated and averaged at the PS to update the global model. The fact that data never leaves the devices is considered to protect its privacy. However, 
recent works \cite{melis2019exploiting, zhu2019deep, geiping2020inverting} show that model updates
or gradients also leak a lot of information about the dataset.
This calls for additional mechanisms to guarantee privacy. Differential privacy (DP) is a widely-adopted rigorous notion of privacy \cite{dwork2014algorithmic}.
Given an algorithm whose output is some statistics about a dataset, if the change in the output probability
distribution is tolerably small, when the input database is changed with a very close neighbouring one, then
the algorithm is deemed differentially private. Many recent works exploit DP-based algorithms to provide rigorous privacy guarantees in machine learning \cite{erlingsson2014rappor,  wang2019collecting, wang2019local, abadi2016deep, mcmahan2017learning, geyer2017differentially}.

We consider co-located devices communicating with the PS over a wireless multiple access channel (MAC). Recent work has shown that rather than the conventional digital communication, the devices can transmit their local updates simultaneously in an uncoded fashion to enable over-the-air computation (OAC) \cite{amiri2020machine, amiri2020federated, zhu2019broadband}. This results in a much more efficient use of the available resources, and significantly improves the learning performance. It is also remarked in \cite{amiri2020federated} that the analog computation in a wireless domain makes privacy more easily attainable. A common method to provide DP guarantees is adding noise to the output with a
variance proportional to the maximum change in the output under neighbouring datasets \cite{dwork2014algorithmic}. In the digital implementation of FL, where each device separately communicates with the PS,
noise is added to the gradients after each local iteration \cite{abadi2016deep} to ensure DP; whereas in analog computation,
the PS only sees the sum of the updates together with the channel noise, which effectively
protects all the updates. Thus, the same amount of protection can be achieved with less perturbation, improving also the final accuracy. 

Privacy in wireless FL through OAC has been recently studied in several works \cite{sonee2020efficient, seif2020, Koda2020DifferentiallyPA, Liu2020privacy}.
In \cite{sonee2020efficient}, distributed stochastic gradient descent (SGD)
is studied with quantized gradients and privacy constraints. In \cite{seif2020}, if the channel noise
is not sufficient to satisfy the DP target, a subset of the devices add additional noise to their updates, benefiting all the devices. 
In \cite{Koda2020DifferentiallyPA} and \cite{Liu2020privacy}, transmit power is adjusted for the same privacy guarantee. We note that, in \cite{seif2020, Liu2020privacy, Koda2020DifferentiallyPA}, channel state information (CSI) at the devices is crucial not only to align their computations at the PS but also for the privacy guarantee. However, in practice, devices acquire CSI from pilots transmitted by the PS. Hence, as an adversary, the PS can adjust the pilots to degrade the privacy level. 
Additionally, in \cite{seif2020}, devices depend on others for privacy, which introduces additional point-of-failure.
 
Differently from the works cited above, we provide privacy in wireless FL by exploiting the anonymity of the transmitters in OAC. Our main contributions can be summarized as follows: (1) We study the effects of randomly sampling the devices participating in each iteration, and batching in local datasets on the privacy. By forcing a constant receive power at the PS across iterations, we provide anonymity to the transmitting devices, and employ it for privacy; (2) By distributing the noise generation
process across the workers, we make the privacy guarantee resilient against the failure of transmitting nodes; (3) Our protocol is robust against pilot attacks.

\section{Preliminaries}

Before presenting the system model and the proposed solution, we give some preliminaries on DP and the threat model.

\subsection{Differential Privacy (DP)}

\begin{defn}
$(\epsilon,\delta)$-DP \cite{dwork2014algorithmic}:
A randomized mechanism ${\cal M}:{\cal D}\to{\cal R}$ is $(\epsilon,\delta)$-differentially
private (DP) if 
\[
\Pr({\cal M}(D)\in{\cal S})\leq e^{\epsilon}\Pr({\cal M}(D')\in \cal S)+\delta,
\]
 $\forall S\subseteq{\cal R}$ and $\forall D,D'\in{\cal D}$ such that $\left\Vert D-D'\right\Vert _{1}\leq1$, which is defined, in this work, as the two very close datasets with the same cardinality differing only in one element. $(\epsilon,\delta)$-DP is a characterization of the change in the output probability distribution of $\cal M$ under small changes in the input dataset. $\epsilon$ characterizes the privacy loss under such a small change and $\delta$ is the upper bound on the failure probability of the bound $\Pr({\cal M}(D)\in{\cal S})\leq e^{\epsilon}\Pr({\cal M}(D')\in \cal S)$. Thus, small $\epsilon$ and $\delta$ are desired.
\end{defn}
\begin{defn}
$(\alpha,\epsilon)$-Rényi DP \cite{mironov2017renyi}:
Rényi divergence between two probability distributions is defined as
\[
D_{\alpha}(P||Q)\triangleq\frac{1}{\alpha-1}\log\mathbb{E}_{x\sim Q(x)}\left(\frac{P(x)}{Q(x)}\right)^{\alpha}.
\]
A randomized mechanism ${\cal M}:{\cal D}\to{\cal R}$ satisfies $(\alpha,\epsilon)$-Rényi
DP (RDP) if $D_{\alpha}\left(\Pr\left({\cal M}(D)=x\right)||\Pr\left({\cal M}(D')=x \right)\right)\leq\epsilon$,
where $\alpha\in[1,\infty)$, $\forall D,D'\in{\cal D}$ such that
$\left\Vert D-D'\right\Vert _{1}\leq1$. $D_\alpha(\cdot||\cdot)$ for $\alpha=1$
and $\alpha=\infty$ are defined by continuity. 
\end{defn}
\begin{defn}
\label{def:Gaussian-Mechanism}
Gaussian Mechanism (GM): A mechanism $\cal M$, which alters the output of another algorithm $f:{\cal D}\to{\cal R}$, by adding Gaussian noise, i.e., ${\cal M}(D)\triangleq f(D)+{\cal N}(0,\sigma^{2})$ is called a GM. GM satisfies $(\epsilon,\delta)$-DP \cite{dwork2014algorithmic} with 
$\epsilon=\sqrt{2\log(1.25/\delta)}\Delta f/\sigma$ and $\Delta f\triangleq\max_{D,D'\in \cal D}\left\Vert f(D)-f(D')\right\Vert _{2}$. It also satisfies $(\alpha,\epsilon')$-RDP with $\epsilon'=\alpha(\Delta f)^{2}/(2\sigma^{2})$ \cite{mironov2017renyi}. 
\end{defn}
\begin{lem}
\label{lem:rdp-to-dp}RDP to DP \cite{mironov2017renyi}: A mechanism
satisfying $(\alpha,\epsilon)$-RDP also satisfies $(\epsilon+\log(1/\delta)/(\alpha-1),\delta)$-DP.
\end{lem}
\begin{lem}
\label{lem:Composition-of-GM}Composition of GM under RDP \cite{mironov2017renyi}:
Composition of $T$ RDP mechanisms operating on the same dataset with
common $\alpha$ and with $(\epsilon_{1},\epsilon_{2},\dots,\epsilon_{T})$
gives an $(\alpha,\sum_{i=1}^{T}\epsilon_{i})$-RDP mechanism.
\end{lem}
As seen in \prettyref{lem:Composition-of-GM}, composition under Rényi
DP is very simple to compute for GM. Moreover, it
gives a much tighter bound on the final $\epsilon$ value compared to
the advanced composition theorem \cite{dwork2014algorithmic}, which
is used in \cite{seif2020, Liu2020privacy}.

\subsection{Threat Model}

We assume that the PS is semi-honest and curious. It is honest in the sense that it wants to acquire the final trained model, and thus, follows the model averaging step accurately, but it also would like to acquire all possible sensitive information about single individual data points. Therefore, it can attack the CSI estimation process
to encourage the devices to increase their transmit
power or to add less noise to their transmissions by suggesting that their channel quality is worse than it is in reality. We note that, if a common pilot signal is used to learn the CSI by all the devices, the CSI values can only be scaled by the same parameter across the devices. 

We assume that the devices are honest and trusted. They do not intentionally attack the learning process; however, we also do not depend on a few devices for privacy as the devices may fail to transmit unintentionally. We further assume that the PS has a single receive antenna, thus it is not capable of determining the identities of the transmitting
devices through transmission directions.

\section{System Model}

We consider $N$ wireless devices. Each device $i\in[N]\triangleq \{1,2,\dots,N\}$ hosts their local dataset
$D_{i}$, which are i.i.d. across the devices. A PS orchestrates these devices to learn a global model $\boldsymbol{w}$ by minimizing a global loss function $\mathcal{L}=\frac{1}{N}\sum_{i\in[N]} \frac{1}{|D_i|} \sum_{d \in D_{i}}  \ell(\boldsymbol{w},d),$ where $\ell(\boldsymbol{w}, d)$ is an application-specific empirical loss function. 
We assume that the dataset is distributed among the devices with the
identical probability and independent of each other. We employ distributed SGD to iteratively minimize $\mathcal{L}$ across the devices. At the beginning of iteration $t$, the PS broadcasts the global model $\boldsymbol{w}_{t}$ to all the devices. Then, the devices participating in the learning process in that iteration transmit their gradient estimates over a Gaussian MAC. In round $t$, the PS receives the superposition of the signals transmitted
by the participating devices as 
\[
\boldsymbol{y}[t]=\sum_{i\in  A_{t}} c_{i,t} \boldsymbol{x}_{i}[t] + \boldsymbol{z}[t],
\]
where $\boldsymbol{z}[t]\sim \mathcal{N}(0,N_0\boldsymbol{I})$ is the channel noise, $A_{t}$ is the set of participating devices, $\boldsymbol{x}_{i}[t]$ is the signal transmitted by device $i$, and $c_{i,t}$ is the channel coefficient from device $i$ to the PS. We assume that the transmitters perfectly know and correct the phase shift in their channels. Therefore, for simplicity, we assume real channel gains in the rest of the paper, i.e., $c_{i,t}\in \mathbb{R^+}$.  

We assume that $c_{i,t}$ is independent across the users and rounds, but remains constant within one round. We further assume that only the agents with sufficiently good channel coefficients participate in each round to increase power efficiency. We assume this happens with probability $p\in(0,1)$ for each device, independently of other devices. We leave the analysis of this probability for different fading distributions as future work. If the device $i$
participates in a round, then it samples uniformly from its own local
dataset such that every data point is sampled independently from the
other samples with probability $q\in(0,1)$. Let $B_{i,t}$
denote the set of samples in the batch used by the $i$th device if it is participating in round $t$. Let $a_t \triangleq |A_t|$ and $b_{i,t}\triangleq|B_{i,t}|$. Both $a_t$ and $b_{i,t}$ are
binomial random variables with parameters $p$ and $q$.

Each participating device computes a gradient estimate based on the random samples in its batch. We bound $L_{2}$-norm of every \emph{per-sample} gradient by $L$. If this is not the case originally, then the gradient is downscaled
by a factor of $\frac{L}{\left\Vert \nabla \ell(\boldsymbol{w}_{t},j)\right\Vert _{2}}$; that is, for a sample $d$, we define $\boldsymbol{g}_{t}(d)\triangleq\nabla \ell(\boldsymbol{w}_{t},d)\times\max\left\{1,\frac{L}{\left\Vert \nabla \ell(\boldsymbol{w}_{t},d)\right\Vert _{2}}\right\}$, as done in \cite{abadi2016deep}.

Device $i$ transmits 
\begin{equation}
\boldsymbol{x}_{i}[t]=h_{i,t}\left(\xi_{i,t}\sum_{j\in B_{i,t}}\boldsymbol{g}_{t}(j)+\beta_{i,t}\boldsymbol{n}_{i,t}\right), \label{eq:grad_update}
\end{equation}
where $\boldsymbol{n}_{i,t}\sim{\cal N}(0,\sigma_{i,t}^{2}\boldsymbol{I})$, $\xi_{i,t}$ and $\beta_{i,t}$ are scaling factors, and $h_{i,t}\triangleq(\tilde{c}_{i,t})^{-1}$.
$\tilde{c}_{i,t}$ is the CSI which can be manipulated by the PS by a
multiplicative factor of $k\in(0,1]$, i.e., $\tilde{c}_{i,t}=k \cdot c_{i,t}$. For simplicity, we assume that the
local datasets are of the same size, although our analysis can be extended easily
by adding another scaling factor to the gradient, if this is not the
case.

The PS updates the model as $\boldsymbol{w}_{t+1}=\boldsymbol{w}_{t}-\eta \boldsymbol{y}[t]$, where $\eta$
is the learning rate. We assume that the number of participating devices
$a_t$ and the batch sizes $b_{i,t}$ are known
to the devices but not to the PS. This can be achieved by keeping
a common random number generator state across the devices, or alternatively
by encrypted communication between the devices. Since this is only sharing two real numbers, the communication overhead is limited.

\section{Main Results}

As we discussed in the previous section, we sample the participating devices and the
local datasets to be used in each iteration. It is well known that the privacy of randomized mechanisms can be amplified by subsampling \cite{balle2018privacy, wang2019subsampled, mironov2019r}; however, these results are for the centralized setting. To take advantage of sampling in the distributed setting, we make sure that all of the data samples
are chosen independently and uniformly with probability $pq$. However,
the challenge is that since the local datasets of different devices
are distinct, the conditional probability of a data point being chosen
given some other data point is already sampled may not be the same as
the marginal probability of the same event. For example, if these
two data points are hosted by the same device, then the conditional probability
of the second data point being chosen is $q$, instead of $pq$.
One way of overcoming this problem is shuffling the whole dataset
across the devices after each iteration to cancel the effect
of locality. However, this would incur a large communication cost. A better
way is to exploit the wireless nature of the protocol as we explain
next. 

\subsection{Ensuring the anonymity of devices} 
Since the PS can only
see the aggregated signal, it cannot know which devices have transmitted. Still, if the PS knows the number of participating devices, then it can collect some information across the iterations and infer some dependency between the samples, such as two samples being hosted by the same device. In such cases, the i.i.d. assumption on the sampling probability of
data points does not hold. Next, we show how we mask the number of participating devices. 

\begin{lem}
\label{lem:sampling_result}
If $\xi_{i,t}=1/b_t,\forall i\in A_{t}$, where $b_t\triangleq \sum_{i\in A_t}b_{i,t}$, the PS cannot infer the number of devices actively transmitting in round $t$.
\end{lem}

\begin{proof}
Since the devices scale their local gradients by $b_t$ before transmitting, and the PS can only see the average gradient at each round, the received power level at the PS is independent of the number or identity of the transmitting devices. Thus from the power level, it is not possible to infer any information about the number of transmitting devices.
\end{proof}

Due to \prettyref{lem:sampling_result}, learning any set of sampled data points does not convey any information to the PS about the remaining sampled data points unless it learns the identities of the transmitting devices. Therefore, batching local samples independently 
with probability $q$ and similarly sampling the transmitting devices independently with probability
$p$ is equivalent to subsampling in the centralized setting
with probability $pq$.

\subsection{Robustness to transmission failures } 

As we have stated, we want our scheme to be robust against transmission failures of devices that are scheduled to transmit at a certain iteration, but failed to do so for some reason. This is achieved by distributing
the noise generation across the devices. 

\begin{lem}
\label{lem:xmit_fail}
If we choose $\beta_{i,t}=1/\sqrt{a_t}$ and $\sigma_{i,t}=\hat{\sigma}_t,\forall i\in A_{t}$, then the received signal at the PS becomes 
\begin{equation}\label{eq:received_signal}
\boldsymbol{y}[t]=\sum_{i\in A_{t}}\frac{1}{b_{t}}\sum_{j\in B_{i,t}}\boldsymbol{g}_{t}(j)+\boldsymbol{n}[t]+\boldsymbol{z}[t], 
\end{equation}
where $\boldsymbol{n}[t]\sim{\cal N}(0,\sigma_t^{2}\boldsymbol{I})$, and $\sigma_t = \hat{\sigma}_t$. When $k<a_t$ devices in $A_t$ fail to transmit, the noise variance degrades to $\sigma_t=\hat{\sigma}_t\sqrt{(a_t-k)/a_t}$.
\end{lem}

\begin{proof}
If $k$ devices fail, there are $a_t-k$ remaining devices transmitting. Since the noise added by the devices are independent and each has the variance $\hat{\sigma}_t^2/a_t$, we have $\sigma_t^2=\hat{\sigma}_t^2(a_t-k)/a_t$. If there is no failure, $k=0$, and we obtain $\sigma_t=\hat{\sigma}_t$.
\end{proof}

As we see in \prettyref{lem:xmit_fail}, in case of transmission failures in $A_t$, the variance of the total noise degrades gracefully, and so does the privacy.


\subsection{Robustness to manipulated CSI values} 

In practice, given a channel noise variance $N_0$, we need to tune $\sigma$ so that the total noise variance $N_0+\sigma ^2$ is sufficient to meet the desired privacy level $(\epsilon,\delta)$. However, if we rely on the channel noise and our channel is better than we think, then the total noise variance may be less than it should be, resulting in larger $\epsilon$ and $\delta$ values than the desired privacy level. This may
cause a privacy breach. To avoid this, in the privacy computations, we simply ignore the channel noise since CSI values are prone to attacks by the PS. Then, the $\epsilon$ value we get is the upper bound of the real $\epsilon$ value in the presence of channel noise.

\subsection{Privacy analysis}

In the next theorem, we present a result which shows the boosting effect of
sampling on the overall privacy.
\begin{thm} \label{thm:RDP-of-main-mech}
Each round $t$ in our FL scheme with OAC 
is $(\alpha,\frac{2p^{2}q^{2}\alpha}{\tilde{\sigma}_t^{2}})$-RDP,  where $\tilde{\sigma}_t\triangleq\frac{\sigma_t b_t}{2L}$,
if $pq\leq1/5$, $\tilde{\sigma_t}\geq4$ and the following inequalities are satisfied 
\begin{equation}
1<\alpha\leq\frac{1}{2}\tilde{\sigma}_t^{2}\log\left( 1+\frac{1}{pq(\alpha-1)} \right)-2\log(\tilde{\sigma}_t),\label{eq:cond1}
\end{equation}
and
\begin{equation}
\alpha\leq\frac{\frac{1}{2}\tilde{\sigma}_t^{2}\log^{2}(1+\frac{1}{pq(\alpha-1)})-\log5-2\log(\tilde{\sigma}_t)}{\log\left(1+\frac{1}{pq(\alpha-1)}\right)+\log(pq\alpha)+ \frac{1}{2\tilde{\sigma}_t^{2}} }.\label{eq:cond2}
\end{equation}
\end{thm}

\begin{proof}
Consider \prettyref{eq:received_signal} and define $\boldsymbol{f}[t]\triangleq \boldsymbol{y}[t]-\boldsymbol{n}[t]-\boldsymbol{z}[t]=\frac{1}{b_t}\sum_{i\in A_{t}}\sum_{j\in B_{i,t}}\boldsymbol{g}_{t}(j)$.
Remember that $\Delta f=\max_{D,D'}$ $\left\Vert f(D)-f(D')\right\Vert _{2}$, where $D$ and $D'$ are two datasets with the same size, differing
only in one data point. For the same realization of random batching and device selection processes, $B_{i,t}(D)$
and $B_{i,t}(D')$ also differ at most in one element. Since $\left\Vert \boldsymbol{g}_{t}(j)\right\Vert _{2}\leq L$,
we find $\Delta f=\frac{2L}{b_t}$. With GM, according to Definition \prettyref{def:Gaussian-Mechanism}, our mechanism is $(\alpha,\alpha (\Delta f)^2/(2\sigma_t ^2))$-RDP without sampling. Moreover, by \prettyref{lem:sampling_result}, the sampling probability of each
sample is $pq$. The remaining of the lemma follows from direct application of \cite[Theorem 11]{mironov2019r}.
\end{proof}
If the conditions in \prettyref{thm:RDP-of-main-mech} are not satisfied,
the RDP of the sampled Gaussian mechanism can still be computed numerically
and improvement due to sampling is still valid. The numerical
procedure for this is given in \cite[Section 3.3]{mironov2019r} and
it is demonstrated in Section \ref{sec:num-res}.

\textbf{Process to compute $(\epsilon,\delta)$-DP after $T$ iterations:} According to \prettyref{lem:Composition-of-GM}, if the received message at round $t$ is
$(\alpha,\epsilon_t)$-RDP, then after $T$ iterations, the mechanism is
$\left(\alpha,\sum_{t\in[T]}\epsilon_t\right)$-RDP. According to \prettyref{lem:rdp-to-dp},
it is $(\sum_{t\in[T]}\epsilon_t+\log(1/\delta)/(\alpha-1),\delta)$-DP. We
observe that although $\epsilon_t$ depends on the parameters in our
mechanism, $\alpha$ does not. It is a parameter we choose to minimize
$(\sum_{t\in[T]}\epsilon_t+\log(1/\delta)/(\alpha-1),\delta)$, which we compute for several $\alpha$ values and take the best $\alpha$ among them.
Since both the analytical (if it exists) and the numerical computations of the composed $(\alpha,\epsilon)$-RDP is computationally cheap, this is feasible.

\section{Numerical Results}\label{sec:num-res}

In this section, we numerically calculate the composite $(\epsilon,\delta)$-DP
of a learning task with $\delta=10^{-5}$, for different sampling rates
given by $pq$, where $pq=1$ means there is no sampling,
i.e., all the devices participate and they use all their local datasets. We assume $\Delta f=1$ and $\sigma_t=1,\forall t\in[T]$. For $pq=1,$ we compute the total privacy spent, i.e., the composite
$\epsilon$ by using both the advanced composition theorem \cite[Theorem 3.20]{dwork2014algorithmic},
which is denoted by `act' in the figure, and the RDP approach
given in \prettyref{lem:Composition-of-GM}. Since we numerically
verify that RDP accounting performs much better for the same sampling rate, we use it for the other values of $p \cdot q$. For numerically
computing the composition with RDP, we used Opacus library \footnote{https://github.com/pytorch/opacus}. We tested $\alpha$ values from
1 to 64, and picked the one minimizing the composite $\epsilon$ value.

\begin{figure}
\centering
\input{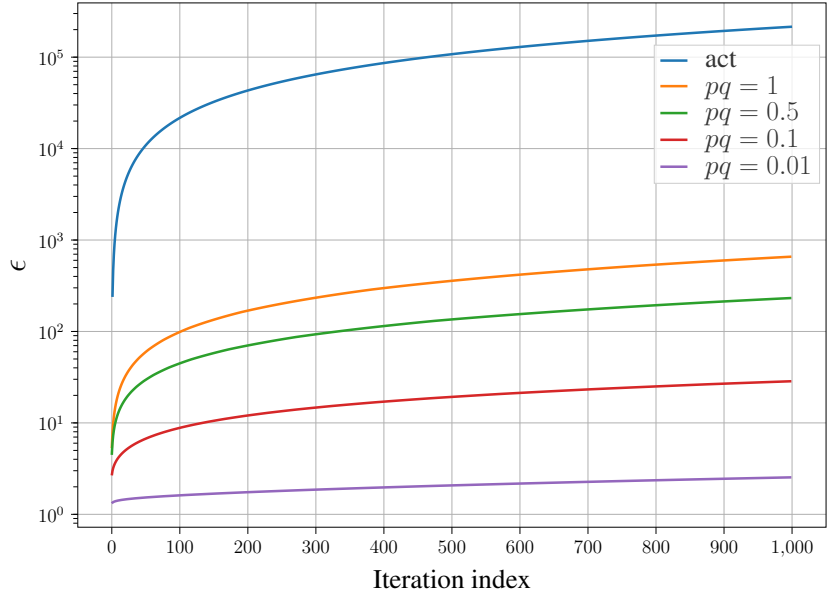}
\caption{Total privacy measured by $\epsilon$ for different sampling rates across iterations.}\label{fig:privacy_comparison}
\end{figure}

We observe in Fig. \ref{fig:privacy_comparison} that for $pq=1$ and $pq=0.5$, the resultant $\epsilon$ values are very high, namely more than 100. This is, in fact, very
weak privacy. Roughly speaking, it means that the output distribution
of the final learned model can change by a factor of 100 when only
one element is changed in the dataset. However, we see that when we have smaller sampling rates, the privacy
guarantee comes into an acceptable range. In an edge network setting, we expect many devices to participate in learning, with substantial local datasets. Therefore, $pq$ values on the order of $0.01$ are not unreasonable. 

\section{Conclusion}

We have exploited the anonymity in wireless transmissions to enable private wireless FL across edge devices. This is achieved by exploiting OAC rather than orthogonal transmissions, which would reveal the identity of the devices. In particular, we exploit random subsampling of both the devices and the local data samples in achieving reasonable DP guarantees. As opposed to recent works on the topic, we do not depend on the channel gain or the noise at the PS for privacy guarantees as these values are prone to attacks; although our method is orthogonal to these techniques and can be combined with them. Our results demonstrate yet another favourable property of OAC in wireless edge learning. We will evaluate the benefits of the proposed privacy mechanism and its comparison with other alternatives on a practical learning problem in our future work.  

\newpage{}

 \bibliographystyle{IEEEbib}
\bibliography{refs}

\begin{thebibliography}{10}

\bibitem{mcmahan2017communication}
Brendan McMahan, Eider Moore, Daniel Ramage, Seth Hampson, and Blaise~Aguera
  y~Arcas,
\newblock ``Communication-efficient learning of deep networks from
  decentralized data,''
\newblock in {\em Artificial Intelligence and Statistics}. PMLR, 2017, pp.
  1273--1282.

\bibitem{melis2019exploiting}
Luca Melis, Congzheng Song, Emiliano De~Cristofaro, and Vitaly Shmatikov,
\newblock ``Exploiting unintended feature leakage in collaborative learning,''
\newblock in {\em 2019 IEEE Symposium on Security and Privacy (SP)}. IEEE,
  2019, pp. 691--706.

\bibitem{zhu2019deep}
Ligeng Zhu, Zhijian Liu, and Song Han,
\newblock ``Deep leakage from gradients,''
\newblock in {\em Advances in Neural Information Processing Systems}, 2019, pp.
  14774--14784.

\bibitem{geiping2020inverting}
Jonas Geiping, Hartmut Bauermeister, Hannah Dr{\"o}ge, and Michael Moeller,
\newblock ``Inverting gradients--how easy is it to break privacy in federated
  learning?,''
\newblock {\em arXiv preprint arXiv:2003.14053}, 2020.

\bibitem{dwork2014algorithmic}
Cynthia Dwork, Aaron Roth, et~al.,
\newblock ``The algorithmic foundations of differential privacy.,''
\newblock {\em Foundations and Trends in Theoretical Computer Science}, vol. 9,
  no. 3-4, pp. 211--407, 2014.

\bibitem{erlingsson2014rappor}
{\'U}lfar Erlingsson, Vasyl Pihur, and Aleksandra Korolova,
\newblock ``Rappor: Randomized aggregatable privacy-preserving ordinal
  response,''
\newblock in {\em Proceedings of the 2014 ACM SIGSAC conference on computer and
  communications security}, 2014, pp. 1054--1067.

\bibitem{wang2019collecting}
Ning Wang, Xiaokui Xiao, Yin Yang, Jun Zhao, Siu~Cheung Hui, Hyejin Shin,
  Junbum Shin, and Ge~Yu,
\newblock ``Collecting and analyzing multidimensional data with local
  differential privacy,''
\newblock in {\em 2019 IEEE 35th International Conference on Data Engineering
  (ICDE)}. IEEE, 2019, pp. 638--649.

\bibitem{wang2019local}
Shaowei Wang, Liusheng Huang, Yiwen Nie, Xinyuan Zhang, Pengzhan Wang, Hongli
  Xu, and Wei Yang,
\newblock ``Local differential private data aggregation for discrete
  distribution estimation,''
\newblock {\em IEEE Transactions on Parallel and Distributed Systems}, vol. 30,
  no. 9, pp. 2046--2059, 2019.

\bibitem{abadi2016deep}
Martin Abadi, Andy Chu, Ian Goodfellow, H~Brendan McMahan, Ilya Mironov, Kunal
  Talwar, and Li~Zhang,
\newblock ``Deep learning with differential privacy,''
\newblock in {\em Proceedings of the 2016 ACM SIGSAC Conference on Computer and
  Communications Security}, 2016, pp. 308--318.

\bibitem{mcmahan2017learning}
H~Brendan McMahan, Daniel Ramage, Kunal Talwar, and Li~Zhang,
\newblock ``Learning differentially private recurrent language models,''
\newblock {\em arXiv preprint arXiv:1710.06963}, 2017.

\bibitem{geyer2017differentially}
Robin~C Geyer, Tassilo Klein, and Moin Nabi,
\newblock ``Differentially private federated learning: A client level
  perspective,''
\newblock {\em arXiv preprint arXiv:1712.07557}, 2017.

\bibitem{amiri2020machine}
Mohammad~Mohammadi Amiri and Deniz G{\"u}nd{\"u}z,
\newblock ``Machine learning at the wireless edge: Distributed stochastic
  gradient descent over-the-air,''
\newblock {\em IEEE Transactions on Signal Processing}, vol. 68, pp.
  2155--2169, 2020.

\bibitem{amiri2020federated}
Mohammad~Mohammadi Amiri and Deniz G{\"u}nd{\"u}z,
\newblock ``Federated learning over wireless fading channels,''
\newblock {\em IEEE Transactions on Wireless Communications}, vol. 19, no. 5,
  pp. 3546--3557, 2020.

\bibitem{zhu2019broadband}
Guangxu Zhu, Yong Wang, and Kaibin Huang,
\newblock ``Broadband analog aggregation for low-latency federated edge
  learning,''
\newblock {\em IEEE Transactions on Wireless Communications}, vol. 19, no. 1,
  pp. 491--506, 2019.

\bibitem{sonee2020efficient}
Amir Sonee and Stefano Rini,
\newblock ``Efficient federated learning over multiple access channel with
  differential privacy constraints,''
\newblock {\em arXiv preprint arXiv:2005.07776}, 2020.

\bibitem{seif2020}
M.~{Seif}, R.~{Tandon}, and M.~{Li},
\newblock ``Wireless federated learning with local differential privacy,''
\newblock in {\em 2020 IEEE International Symposium on Information Theory
  (ISIT)}, 2020, pp. 2604--2609.

\bibitem{Koda2020DifferentiallyPA}
Yusuke Koda, Koji Yamamoto, Takayuki Nishio, and Masahiro Morikura,
\newblock ``Differentially private aircomp federated learning with power
  adaptation harnessing receiver noise,''
\newblock {\em arXiv preprint arXiv:2004.06337}, 2020.

\bibitem{Liu2020privacy}
Dongzhu Liu and Osvaldo Simeone,
\newblock ``Privacy for free: Wireless federated learning via uncoded
  transmission with adaptive power control,''
\newblock {\em arXiv preprint arXiv:2006.05459}, 2020.

\bibitem{mironov2017renyi}
Ilya Mironov,
\newblock ``R{\'e}nyi differential privacy,''
\newblock in {\em 2017 IEEE 30th Computer Security Foundations Symposium
  (CSF)}. IEEE, 2017, pp. 263--275.

\bibitem{balle2018privacy}
Borja Balle, Gilles Barthe, and Marco Gaboardi,
\newblock ``Privacy amplification by subsampling: Tight analyses via couplings
  and divergences,''
\newblock in {\em Advances in Neural Information Processing Systems}, 2018, pp.
  6277--6287.

\bibitem{wang2019subsampled}
Yu-Xiang Wang, Borja Balle, and Shiva~Prasad Kasiviswanathan,
\newblock ``Subsampled r{\'e}nyi differential privacy and analytical moments
  accountant,''
\newblock in {\em The 22nd International Conference on Artificial Intelligence
  and Statistics}. PMLR, 2019, pp. 1226--1235.

\bibitem{mironov2019r}
Ilya Mironov, Kunal Talwar, and Li~Zhang,
\newblock ``R{\'e}nyi differential privacy of the sampled gaussian mechanism,''
\newblock {\em arXiv preprint arXiv:1908.10530}, 2019.

\end{thebibliography}

\end{document}